\documentclass[reprint, amsmath,amssymb,aps,pra, superscriptaddress, hidelinks, floatfix
]{revtex4-2}

\pdfoutput=1

\usepackage[english]{babel}
\usepackage{graphicx,siunitx,xstring,hyperref}
\usepackage{dcolumn}
\usepackage{bm}
\usepackage{physics,mathrsfs, chemformula}

\usepackage[activate={true,nocompatibility}
            ,final
            ,tracking=true
            ,kerning=true
            ]{microtype}
    \microtypecontext{spacing=nonfrench}

\makeatletter
\renewcommand*{\fnum@figure}{{\normalfont\bfseries \figurename~\thefigure}}
\makeatother

\begin{abstract}
We present a theoretical study of dielectric bowtie cavities and show that they are governed by two essentially different confinement regimes. The first is confinement inside the bulk dielectric and the second is a local lightning-rod regime where the field is locally enhanced at sharp corners and may yield a vanishing mode volume without necessarily enhancing the mode inside the bulk dielectric. We show that while the bulk regime is reminiscent of the confinement in conventional nanocavities, the most commonly used definition of the mode volume gauges in fact the lightning-rod effect when applied to ultra-compact cavities, such as bowties. Distinguishing between these two regimes will be crucial for future research on nanocavities, and our insights show how to obtain strongly enhanced light-matter interaction over large bandwidths.
\end{abstract}

\begin{document}

\title{Two regimes of confinement in photonic nanocavities: bulk confinement versus lightning rods}

\newcommand{\fotonik}{Department of Photonics Engineering, DTU Fotonik, Technical University of Denmark, Building 343, DK-2800 Kgs.\ Lyngby, Denmark}
\newcommand{\nanophoton}{NanoPhoton - Center for Nanophotonics, Technical University of Denmark, Ørsteds Plads 345A, DK-2800 Kgs.\ Lyngby, Denmark}

\author{Marcus Albrechtsen} \email{maralb@fotonik.dtu.dk} \affiliation{\fotonik}
\author{Babak Vosoughi Lahijani} \affiliation{\fotonik} \affiliation{\nanophoton}
\author{S{\o}ren Stobbe} \affiliation{\fotonik} \affiliation{\nanophoton}

\date{\today}
\maketitle

\section{Introduction}

The confinement of light in optical nanocavities \cite{painter_two-dimensional_1999,akahane_high-q_2003,notomi_manipulating_2010} enhances the inherently weak interaction between light and matter \cite{lodahl_interfacing_2015}. This is important for a wide span of applications ranging from optomechanics over lasers to biosensing and emerging quantum technologies \cite{vahala_optical_2003, lodahl_interfacing_2015,koenderink_nanophotonics_2015,ma_progress_2020}. Each application has unique figures of merit \cite{notomi_manipulating_2010,choi_self-similar_2017}, and here we consider enhancement of the decay rate in a nanostructure compared to a homogeneous medium, i.e., the ratio of the local density of optical states (LDOS) to the density of optical states, which is commonly denoted the Purcell factor \cite{purcell_spontaneous_1946},
\begin{equation}
    \label{eq:purcell}
    P = \frac{3}{4\pi^2} \left(\frac{\lambda}{n}\right)^3 \frac{Q}{V}.
\end{equation}
Here $\lambda$ is the resonant wavelength of the nanocavity, $n$ is the refractive index, and $V$ is effective mode volume. Notably, both $n$ and $V$ depend on position and must be evaluated at the position of the emitter, $\mathbf{r}_0$. The quality factor, $Q$, describes the localization in frequency space of the cavity mode due to resonant features.
Most previous research on dielectric cavities focused on enhancing the Purcell factor through high quality factors \cite{akahane_high-q_2003,kippenberg_demonstration_2004} since it was believed that the mode volume was bounded at the diffraction limit \cite{coccioli_smallest_1998,deotare_high_2009,khurgin_how_2015}, $V_\text{DL}=[\lambda/(2n)]^3$. This has provided important advances in light-matter interaction during the past decades, but high-$Q$ cavities are unsuited for broadband interactions with pulses required for, e.g., electronics-photonics integration \cite{mork_squeezing_2020}. Additionally, it has become increasingly difficult to further increase $Q$ in nanostructures due to structural disorder and other ubiquitous surface effects \cite{minkov_automated_2014,sekoguchi_photonic_2014}. Plasmonic cavities can realize deep subwavelength $V$, but are intrinsically limited by ohmic losses to $Q<100$ \cite{wang_general_2006}. These considerations call for research into decreasing the mode volume of dielectric nanocavities.

Dielectric cavities with mode volumes below the diffraction limit can be realized using bowtie structures. These were first identified by Gondarenko et al.\ \cite{gondarenko_spontaneous_2006} who used an inverse-design algorithm to produce a structure from which the authors intuited resonant features that define a single cavity mode with a bowtie in the center and mode volumes below the diffraction limit. In later work, Gondarenko et al.\ combined this concept with larger, high-$Q$ cavities to further enhance the $Q/V$-ratio \cite{gondarenko_low_2008}. Later works on bowtie cavities \cite{liang_formulation_2013,hu_design_2016,choi_self-similar_2017,wang_maximizing_2018,hu_experimental_2018,mignuzzi_nanoscale_2019,zhou_ultra-low_2019,zhao_minimum_2020} expanded on these concepts.
Our work is motivated by this line of research where in several cases mode volumes below the diffraction limit are claimed even for modes extending over regions much larger than the $\simeq(\SI{200}{\nano\meter})^3$ volume corresponding to the diffraction limit at telecom wavelengths in silicon.
This apparent yet so far overlooked inconsistency calls for a detailed study of the mode volume and the mechanisms for photon confinement in nanocavities. This is what we report here and we find that two related yet very different effects govern the confinement: Lightning-rod effects, which appear due to localized surface fields at sharp features, and bulk confinement, which confines light inside the dielectric material. We present detailed numerical investigations of the two effects, which point to several new insights that were so far overlooked. For example, we observe that lightning-rod effects are highly localized and barely change the global structure of the mode, which explains how subdiffraction mode volumes can appear in spatially extended modes. We further show that the radius of curvature plays a crucial role, which, if neglected, can result in arbitrarily small but unphysical mode volumes regardless of whether or not the mode is globally confined.

\section{Theory}

\begin{figure*}[ht]
    \centering
    \includegraphics[width=0.88\linewidth]{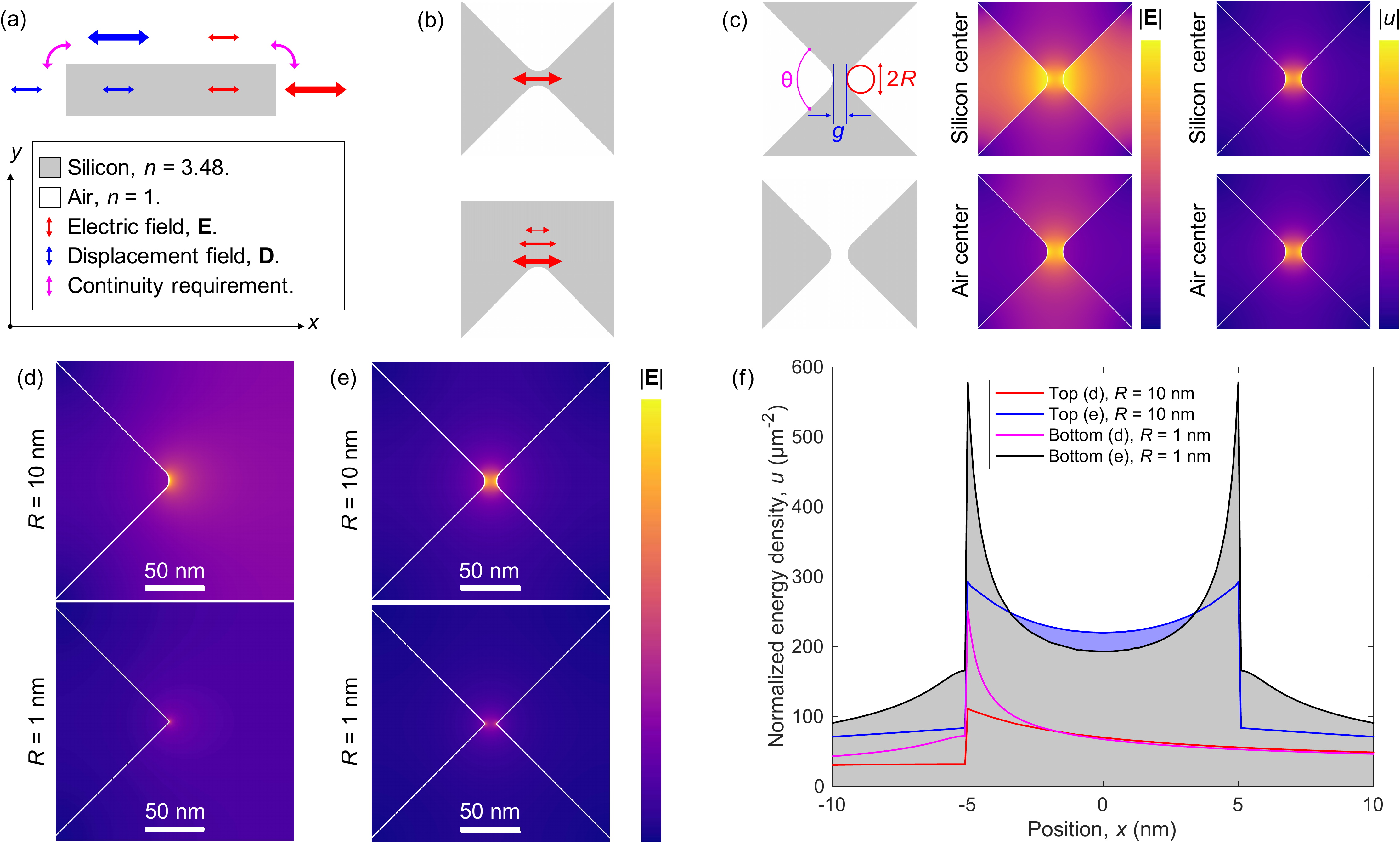}
    \caption{
    Working principles of a dielectric bowtie.
    (a), Boundary effects on the electric field (red arrow), \textbf{E}, and the displacement field (blue arrow), $\mathbf{D} = \epsilon\mathbf{E}$, across dielectric interfaces. Larger arrows imply enhancement of the associated field and the magenta arrows illustrate the continuity requirement of the potential around corners, causing lightning-rod effects and field singularities if the corners are sharp.
    (b), Two oppositely facing lightning rods form a bowtie (top) where the field is enhanced between them. In contrast, the field decays rapidly away from a single lightning rod (bottom).
    (c), Bowties defined by a gap, $g = \SI{20}{nm}$, constructed of two facing cones of angle $\theta=\SI{90}{\degree}$, controlled by a radius of curvature of the tip, $R = \SI{20}{nm}$, which enhances the field everywhere between the tips in silicon (air) in the top (bottom) row. The central (last) column shows the electric field amplitude (energy density) calculated as a 2D electrostatic problem using a finite-element method.
    (d-e), Electric energy of a single cone and a bowtie, respectively, each with different $R$.
    (f), Normalized energy density, $u$, along the center of (d-e). All colormaps (c)-(e) are normalized to their maximum.
    }
    \label{fig:1}
\end{figure*}

A resonant mode has a certain distribution of the electric field, $\mathbf{E}$, with an associated energy density, $u_0=\mathbf{D}\cdot\mathbf{E}=\epsilon_0\epsilon_r\mathbf{E}\cdot\mathbf{E}$, and the total energy in the mode is $\int u_0\text{d}\mathbf{r}'$. This defines a normalized energy density with dimensionality of inverse volume,
\begin{equation}
\label{eq:u}
    u(\mathbf{r}) = \frac{
    \epsilon_r(\mathbf{r}) \mathbf{E}(\mathbf{r}) \cdot \mathbf{E}(\mathbf{r})}{
    \int_\mathbf{r'} \epsilon_r(\mathbf{r}')\mathbf{E}(\mathbf{r}')\cdot\mathbf{E}(\mathbf{r}') \text{d}\mathbf{r'}},
\end{equation}
where $\epsilon_r=n^2$ is the dielectric constant, and here we consider silicon with $n=3.48$. In general, $u$ is complex but for high-$Q$ cavities it is an excellent approximation to assume that $u$ is purely real \cite{kristensen_generalized_2012}. To obtain the strongest possible enhancement of spontaneous emission from a dipole emitter in a cavity, the dipole should be placed at the most intense point of the mode in the cavity, which defines the effective mode volume as the inverse of the normalized energy density at that position \cite{kristensen_generalized_2012,lodahl_interfacing_2015,notomi_manipulating_2010}, $V(\mathbf{r}_0)=1/\text{Re}\{u(\mathbf{r}_0)\}$. More generally, the emitter should be placed in the position with the highest LDOS, which is a function of frequency, polarization, and position.

We note that besides enhancement of spontaneous emission, which is quantified by the Purcell factor, cf.\ Eq.~(\ref{eq:purcell}), other applications of light-matter interactions such as two-photon absorption and Kerr nonlinearities are governed by different mode volumes with, e.g., different powers of the field or the volume integral only over regions with nonlinear materials \cite{notomi_manipulating_2010}. For the remainder of this work, we limit the discussion to the Purcell factor and how its mode volume can be reduced using the boundary conditions of Maxwell's equations.

\section{Electrostatic analysis}
\subsection{Bowties versus lightning rods}

Before proceeding with full three-dimensional (3D) electrodynamic calculations, we first consider a two-dimensional (2D) finite-element electrostatic model to gain insight into the scaling of the energy density, and thus mode volume, with cavity dimensions and geometry. The electrostatic model is a valid approximation to the full electrodynamic problem because we consider deep-subwavelength features \cite{khurgin_how_2015,choi_self-similar_2017}. The relevant figure of merit in this model is the mode area, $A(\mathbf{r}_0)$, which appears when the integral over $\text{d}\mathbf{r}'$ is evaluated in 2D \cite{kippenberg_demonstration_2004}.

Figure~\ref{fig:1}(a) shows the boundary conditions across a dielectric material interface, which demand that the tangential component of the electric field, $\textbf{E}$, and the normal component of the displacement field, $\textbf{D}=\epsilon\textbf{E}$, are continuous. The opposite configuration leads to discontinuous fields that differ by a factor of $\epsilon_r$.
The magenta arrows indicate geometrically sharp vertices at which the electric field must diverge locally. Figure~\ref{fig:1}(b) shows two facing tips that are close to each other and form a bowtie \cite{gondarenko_spontaneous_2006,gondarenko_low_2008,liang_formulation_2013,hu_design_2016,choi_self-similar_2017,wang_maximizing_2018,hu_experimental_2018,mignuzzi_nanoscale_2019,zhou_ultra-low_2019,zhao_minimum_2020}, thus enhancing the field between them. A single tip results in a locally enhanced field that falls off rapidly away from it.
The field divergences at sharp tips imply lightning-rod singularities to ensure continuity of the electric potential, which is required by the electromagnetic continuum model of materials \cite{sommerfeld_mathematical_2004,andersen_field_1978,van_bladel_field_1985,landau_electrodynamics_1984,jackson_classical_1999}. The potential remains continuous so the energy is well defined. However, the field, and thus the energy density, locally diverges at such points, which implies that local evaluations of the mode volume based on the point of maximum $u$, is prone to gauge these divergent fields. Since perfectly sharp corners are unphysical, so is the associated mode volume.

\begin{figure*}[ht]
    \centering
    \includegraphics[width=0.88\linewidth]{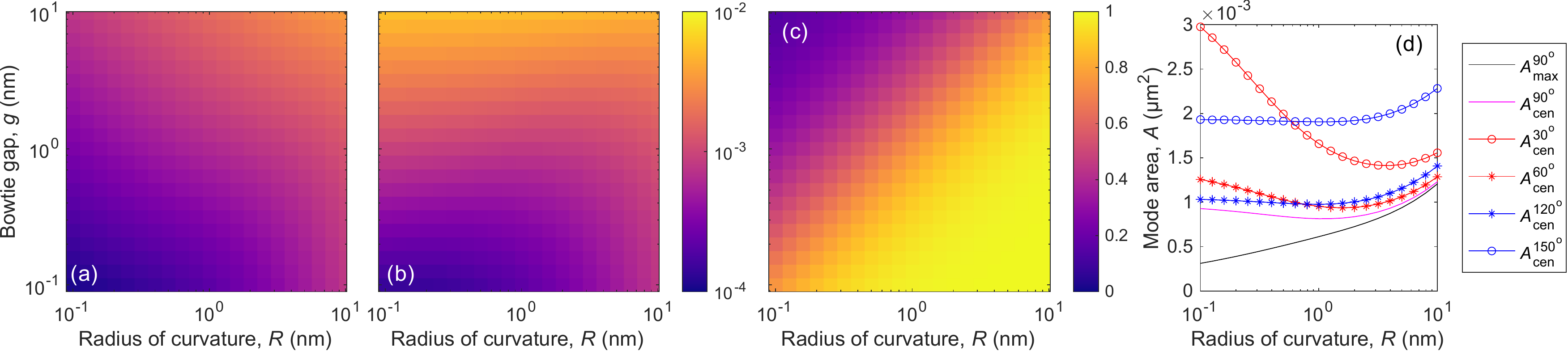}
    \caption{
    Scaling of the mode area with the radius of curvature, $R$, gap, $g$, and bowtie angle, $\theta$.
    (a-b), Scaling of $A$ (in \si{\micro\meter\squared}) evaluated at the maximum (at material boundary), $A_\text{max}$, and at the geometric center, $A_\text{cen}$, respectively.
    (c), Linear ratio between (a) and (b), $A_\text{max}/A_\text{cen}$.
    (d), Mode area $A_\text{max}$ (black line) and $A_\text{cen}$ (magenta line) for \SI{90}{\degree} bowtie angles as a function of $R$, i.e., horizontal cuts through the center of (a-b). Wider angles (blue lines) correspond to the geometry approaching a slot, which does not have a lightning-rod regime but also cannot reduce $A_\text{cen}$ as much as a sharper bowtie. More narrow angles (red lines) exhibit more pronounced differences between the two confinement regimes and has minimal $A_\text{cen}$ for $R>g$ depending on the angle. The smallest mode area is obtained for $\theta=\SI{90}{\degree}$ with $R=g$.
    }
    \label{fig:2}
\end{figure*}

In finite-element methods and other meshed numerical models, the calculated mode volume will depend on the size of the mesh and will not converge. Instead, to ensure converging results that describe the system instead of the mesh, a finite sharpness must be imposed on these lightning-rods, i.e., using a finite radius of curvature, $R$, as shown in Fig.~\ref{fig:1}(c). The mesh must obviously be much smaller than the radius of curvature it aims to resolve. The angle, $\theta$, is \SI{90}{\degree} everywhere in this work unless specified.
Here the field is enhanced between the tips and in the limit where $R\rightarrow\infty$, the bowtie reduces to a slot \cite{almeida_guiding_2004,robinson_ultrasmall_2005,schneider_strong_2016}, and when $R\rightarrow0$, the bowtie becomes two facing lightning rods with diverging fields.
Figure~\ref{fig:1}(c) further compares bowties with solid (void) centers in the top (bottom) row, where the central column shows $|\mathbf{E}|$ and the last column shows the energy density, $u$. Notably, while the electric field appears more localized in the air center compared to the dielectric center, the energy distributions of both bowtie types are identical. For this reason, we will denote the central part between the bowtie tips as a gap, regardless of whether the refractive index of the gap is smaller or larger than that of the tips. For example, our considerations apply to both dielectric gaps (i.e., bridges) surrounded by air tips and an air gap between dielectric tips; only it should be noted that the polarization of the electric field is rotated by \SI{90}{\degree} between the two cases due to the different boundary conditions \cite{choi_self-similar_2017}. These insights show that the radius of curvature plays a decisive role for nanocavities based on boundary effects, which we explore in further detail below.

Figure~\ref{fig:1}(d-e) show $u$ for a lightning-rod and a bowtie respectively, with a gap, $g=\SI{10}{nm}$, and a radius of curvature of the tip in the top (bottom), $R=\SI{10}{nm}$ ($R=\SI{1}{nm}$). Figure~\ref{fig:1}(f) shows the normalized energy density, $u$, evaluated at different positions along the cavity center for the four structures shown in Figs~\ref{fig:1}(d-e).
This figure illustrates a main insight of our work: While the field can be enhanced locally by both lightning rods and bowties, the field maximum is larger but falls off much more rapidly for a lightning rod compared to a bowtie with the same radius of curvature. This means that although a lightning rod creates a region of high field intensity, the field is mostly enhanced at the surface. In addition, the bowtie effect is stronger because the long tail of the lightning rod that would have appeared if the bowtie had had only one boundary is suppressed by the presence of the oppositely facing boundary, and the energy density is normalized to the total energy in the system, cf.\  Eq.~\ref{eq:u}. Another important insight from this figure is that while reducing the radius of curvature for a bowtie enhances the maximum field, which occurs at the material interface, the energy density $u$ in the bulk can be reduced when $R$ decreases. The blue-shaded region highlights the range where this reduction in the light-matter interaction occurs.

The sharpest possible tip induces the smallest mode volume at the interface, which is relevant for light-matter interaction at the very surface. However, this is seldomly desirable due to, e.g., nonradiative surface recombination. In practice, such surface effects are extremely sensitive to, or even indistinguishable from, structural disorder, and besides the extreme care needed in design, nanofabrication, and numerical calculations to approach such effects, they would also raise more fundamental questions about the validity of the electromagnetic continuum model of materials, local field effects, and other complex surface phenomena such as oxidation and condensation \cite{sekoguchi_photonic_2014,asano_photonic_2017}.
Approaching this in a consistent way would be a daunting research challenge that has not been attempted so far.
To summarize the findings of the preceding paragraphs: It has no physical meaning to discuss bowtie or lightning-rod cavities without specifying the radius/radii of curvature. This is in part because a radius of curvature of zero represents an unphysical extrapolation of continuum theory and in part because the mode volume goes to zero when the radius of curvature goes to zero. In addition, the smallest possible mode volume in the center of a bowtie cavity, which is typically the relevant position for experiments and technology, is not obtained for the smallest possible radius of curvature but rather at some optimum value as discussed in further detail below. 

\subsection{Scaling of bowties structures}

The findings of the preceding section call for a systematic study of how the mode area depends on the gap and the radius of curvature. Figure~\ref{fig:2}(a-b) shows the mode area (in \si{\micro\meter\squared}) evaluated at the point of maximum energy density (i.e., at the material interface), $A_\text{max}$, and at the center of the bowtie, $A_\text{cen}$, as a function of $R$ and $g$. We observe that reducing the gap always reduces both mode areas, implying an enhanced light-matter interaction. In contrast, reducing the radius of curvature always reduces $A_\text{max}$ but not always $A_\text{cen}$. This is shown more directly in Fig.~\ref{fig:2}(c) where the ratio $A_\text{max}/A_\text{cen}$ is shown. The diagonal contours along $g=R$ show that the two evaluations can give significantly different results for some parameters while they are equivalent for large $R$ and small $g$. We note that these observations are consistent with Fig.~\ref{fig:1}(f).
Figure~\ref{fig:2}(d) shows the two mode areas as a function of $R$ for a gap of $g=\SI{1}{nm}$, i.e., the figure shows a horizontal scan through Fig.~\ref{fig:2}(a-b). Additionally, $A_\text{cen}$ is shown for multiple other bowtie angles with wider angles corresponding to a more slot-like system and where narrow angles results in lightning rods for larger $R$ compared to the angle, $\theta=\SI{90}{\degree}$. Notably, $A_\text{cen}$ exhibits a minimum at $R_\text{min}=g$ for $\theta=\SI{90}{\degree}$, indicating optimum parameters for bowties confining light inside the material under the electrostatic approximation. We further observe that $A_\text{cen}$ rapidly increases for $R<R_\text{min}$ for narrow-angled bowties, while $A_\text{cen}$ is substantially unaffected for the wide-angled bowties and instead plateaus at a larger mode area compared to the optimal case of $\theta=\SI{90}{\degree}$. This is consistent with the scaling of a slot, which does not need a finite radius of curvature but also cannot achieve as small mode areas as bowtie structures.

We denote the region $R<R_\text{min}$ the lightning-rod regime of bowtie cavities, with $R_\text{min}=g$ in the optimal case of $\theta=\SI{90}{\degree}$: Cavities aiming for interactions at the surface could aim to make the sharpest possible tips to enhance the interaction, while cavities relying on emitters inside the material should aim for $R\sim R_\text{min}$.
These effects can also be relevant for conventional nanocavities, such as H0 cavities, but it is not pronounced in cavities with a global antinode in the very center, i.e., inside the bulk, such as the fundamental mode of an L3 cavity. Notably, it can be important for higher-order modes, which may seem to have a smaller volume than the fundamental mode due to field discontinuities at interfaces \cite{nakayama_effect_2011}.

Cavities approaching the lightning-rod regime must carefully consider the position at which their figures of merit are computed, such as the mode volume.
We observe that lightning rods cause local spikes in the field that do not substantially change the volume integral, and, thus, the overall energy distribution within the mode. In turn, they depend on the local polarization of the field of the dominating cavity mode. In other words, lightning-rod effects are local effects that may imply locally enhanced fields but they do not imply that the mode is globally confined to a small volume. For example, a sharp dielectric defect on the surface of a waveguide, or even just any corner of a rectangular waveguide, would imply a mode volume that can be arbitrarily small (given by the radius of curvature).

We denote the region $R\ge R_\text{min}$ the bulk confinement regime of bowtie cavities in which the field falls off slowly within the gap, which causes an enhanced field throughout the bulk material and not just at the boundaries. This substantially affects $u$, since the volume integral normalizes it, as well as the other cavity figures of merit, i.e., evaluation in the center of the bowtie yields a value corresponding to confinement in the bulk. To summarize these findings: The mode volume evaluated at the center of a bowtie cavity describes confinement in the bulk and resembles the concept of the mode volume commonly discussed in the literature for other types of nanocavities. In contrast, the mode volume evaluated a the field maximum for bowtie cavities describes a local lightning-rod effect at material boundaries. This can be equivalent to $V$ inside the structure if $R\ge R_\text{min}$ but otherwise the two definitions are not comparable.

\section{Electrodynamic analysis}
\subsection{Lightning-rod effects in conventional cavities}

\begin{figure*}
    \centering
    \includegraphics[width=.88\linewidth]{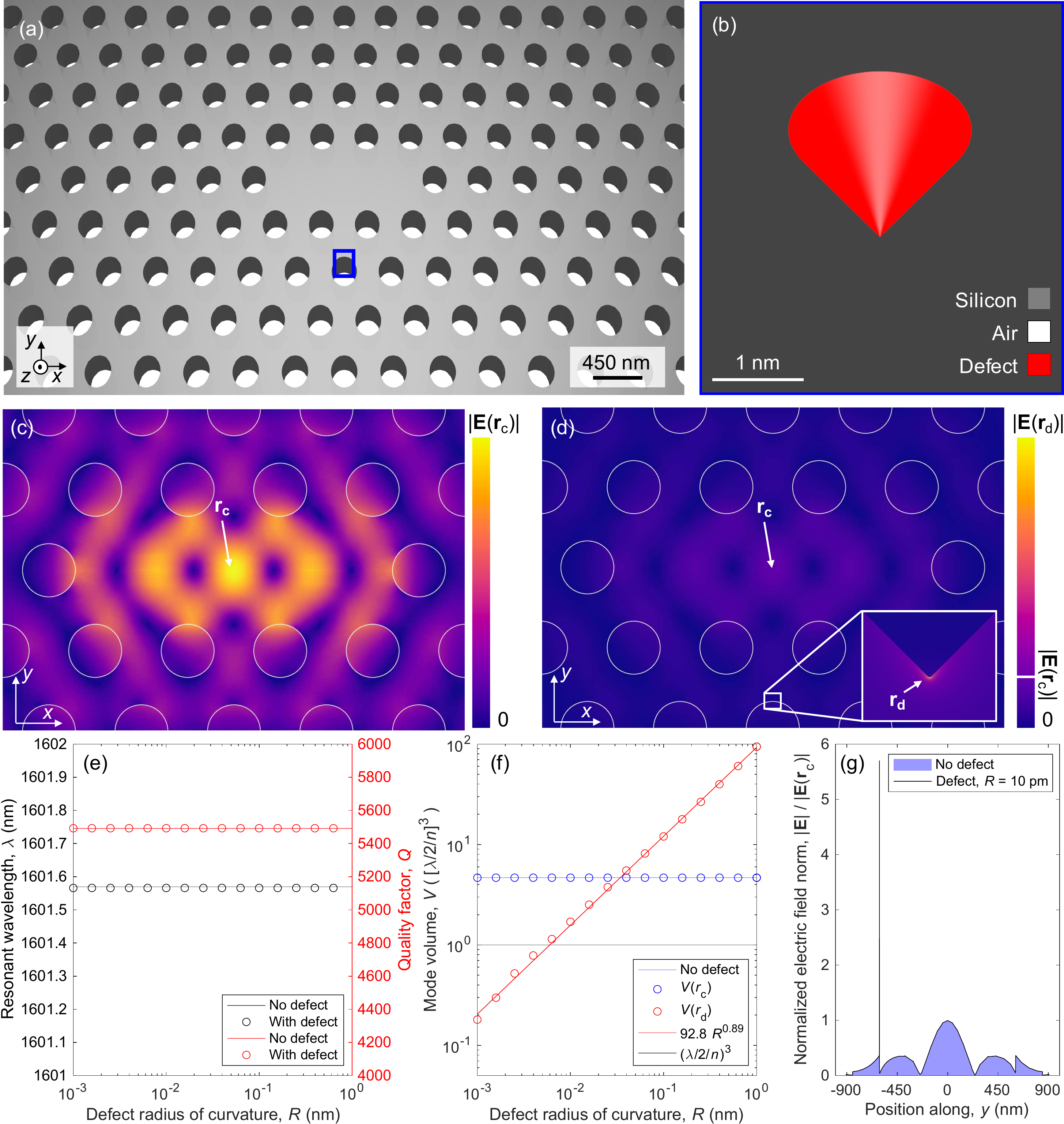}
    \caption{
    Mode volumes below the diffraction limit in an L3 cavity due to lightning-rod effects at a cone far away from the cavity center.
    (a), Illustration of an L3 cavity where 3 holes are removed along a line in the photonic crystal. The blue square shows the position of a defect ($\mathbf{r}_d=(y=a\sqrt{3}-R, x=z=0)$), enlarged in (b).
    (c), The cavity mode calculated without the defect, and (d) with the defect limited by a radius of curvature, $R=\SI{10}{pm}$. The inset in (d) shows the mode around the tip of the defect. Note that the mode profile is identical to (c) everywhere except around the defect but appears different as the colormap is normalized to the field at $\mathbf{r}_d$.
    (e), The resonant wavelength and quality factor for the structure without a defect (solid line) and with a defect (circles) for a number defect radius of curvatures, $R$. Evidently, the defect has no significant effect.
    (f), The mode volume evaluated in the center, $V(\mathbf{r}_c)$, and at the defect tip, $V(\mathbf{r}_d)$. The red line is a power-law fit $V(\textbf{r}_d,R)$ showing an almost linear dependence on $R$ of the defect in this particular case.
    (g), The norm of the electric field, $|\mathbf{E}(y,x=z=0)|$. The black outline is for a defect with $R=\SI{10}{pm}$ and the blue shaded area is the mode without a defect. They are identical everywhere except at $\mathbf{r}_d$ where the field has a local spike.
    }
    \label{fig:3}
\end{figure*}

While 2D electrostatics can provide insights into the scaling of bowties, we will henceforth consider 3D electrodynamic theory to illustrate the impact of lightning-rod effects in cavities studied in the literature. We first investigate an L3 photonic-crystal cavity \cite{akahane_high-q_2003} (with no shifted holes) shown in Fig.~\ref{fig:3}(a). We consider a pitch of $a=\SI{420}{nm}$, a hole radius of $b=\SI{120}{nm}$, and a membrane thickness of $t=\SI{250}{nm}$. The photonic crystal has 9 periods along $x$ and 7 periods along $y$. In addition to the conventional design, we add a nanometer-scale defect to the simulation domain, acting as a lightning-rod. Figure~\ref{fig:3}(b) shows the defect introduced in the side of the cavity shaped as a cone of silicon with a height and base radius $h=\SI{1}{nm}$. The very tip of the defect is rounded to a sphere of radius $R=\SI{10}{pm}$.
Figure~\ref{fig:3}(c-d) shows the eigenmode of the structure without and with the defect, both calculated in 3D using a finite-element method. The mesh at the apex of the defect is much smaller than $R$ such that the model converges with finer mesh, and instead the divergence is controlled directly by $R$. The effect of the tip is directly observed from the normalized field plots (Fig.~\ref{fig:3}(c) and (d)).

To show explicitly that the perturbation by small defects does not affect the mode at the center of the L3 cavity, we first ensured convergence of the unperturbed structure and then kept this mesh constant in the structure \SI{5}{nm} away from the defect to avoid numerical fluctuations and only show the effects of the defect. The small defect acts as a lightning rod, enhancing the field locally at the tip in a vanishing volume, controlled by the finite radius of curvature \cite{andersen_field_1978,van_bladel_field_1985}. This has a negligible effect on the cavity figures of merit as shown in Fig.~\ref{fig:3}(e-f), i.e., the resonant wavelength, the quality factor, and the mode volume at the center of the L3 cavity. This is because the divergent field happens in a vanishing volume, and therefore, the contribution from such a sharp lightning rod to the volume integral fades with mesh refinement. Defects may in general reduce the quality factor but this is not observed due to the modest quality factors of the L3 cavity studied here.
Figure~\ref{fig:3}(g) shows a line-scan through the cavity where the defect is visible as a local enhancement of the electric field on top of the global field intensity of the resonant structure.
Notice that the smallest mode volume is found in air, not silicon, since the defect is a protrusion. It is possible to make a similar defect, which achieves a deep-subwavelength $V$ in the high-index material instead by simply observing the boundary conditions and the local polarization of the field of the L3 cavity mode \cite{andersen_field_1978,van_bladel_field_1985} to make a reentrant defect instead. For example, a reentrant defect at the first hole along $(x, y=z=0)$ results in $V<(\lambda/(2n))^3$ for a radius of curvature $R\le\SI{1}{pm}$.
This study verifies that the mode volume calculated by the maximum evaluation can go to zero when the domain contains sharp features without affecting the global structure of the cavity mode. In practice, however, such features are indiscernible from structural disorder and numerically the divergence will be resolved by the finite mesh size. Importantly, the addition of such features does not change the light-matter interaction strength elsewhere in the cavity including the center.

\subsection{Bulk confinement versus lightning rods in bowtie cavities}

\begin{figure*}
    \centering
    \includegraphics[width=0.88\linewidth]{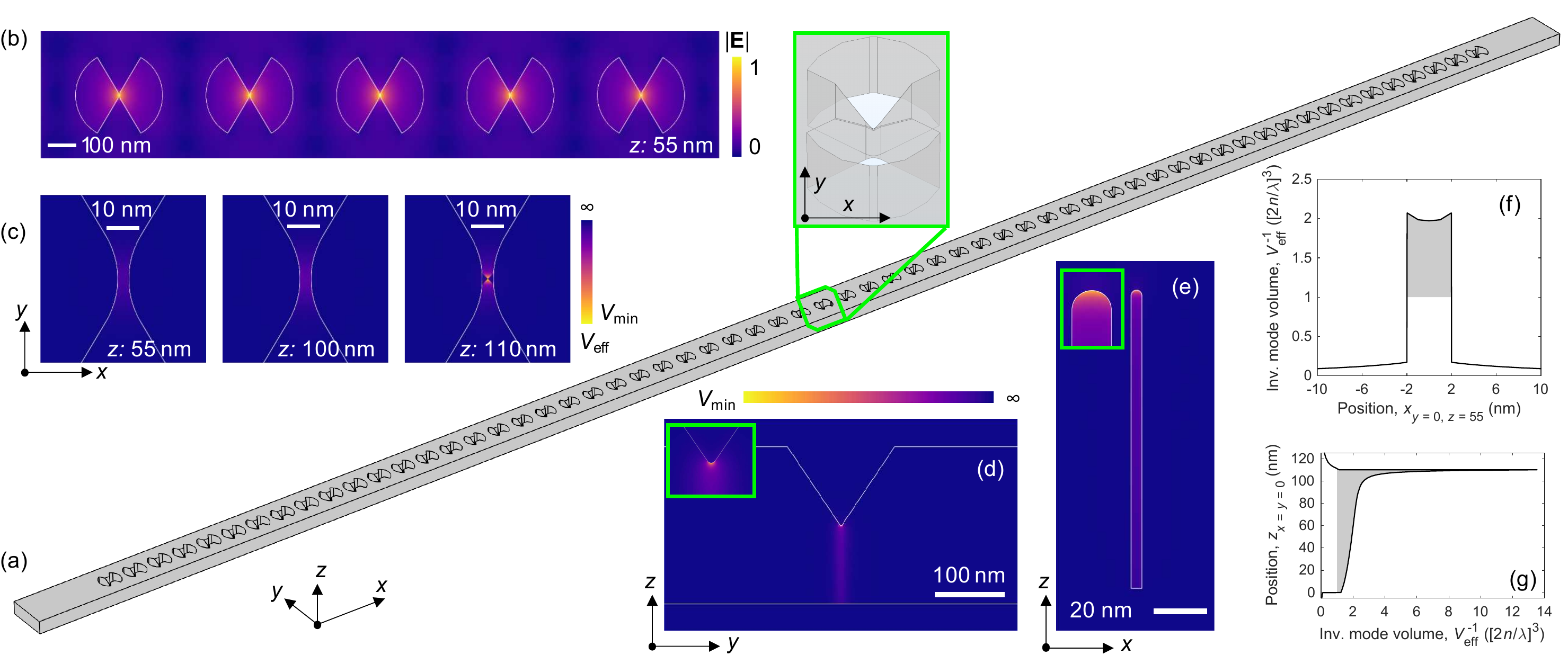}
    \caption{
    Model of a nanocavity with an array of V-groove bowties that have a gap, $g = \SI{4}{nm}$, where lightning-rod effects imply mode volumes deep below the diffraction limit at the V-groove tip and bulk confinement slightly below the diffraction limit.
    (a), The cavity is a suspended photonic-crystal nanobeam-cavity defined in silicon. The inset shows a V-groove penetrating half way into the bowtie that creates a lightning-rod effect.
    (b-c), Electric field norm, $|\mathbf{E}|$, and effective mode volume, $V$, respectively, in the ($xy$) cross-section plane, shown at the center of the structure, close to the interface, and at the interface.
    (d-e), Side-view of $V$ in $yz$- and $xz$-planes with insets showing the apex of the lightning-rod.
    (f-g), Line-scans through the structure showing $1/V$ normalized to the inverse of the diffraction limit; the gray-shaded area highlights the regions with a mode volume below the diffraction limit.
    The cavity is solved as an eigenfrequency problem in 3D with a finite element method using $\sim\SI{900}{GB}$ RAM to resolve the many narrow features with high aspect ratios with a mesh size well below $g$.
    }
    \label{fig:4}
\end{figure*}

\begin{figure*}
    \centering
    \includegraphics[width=0.88\linewidth]{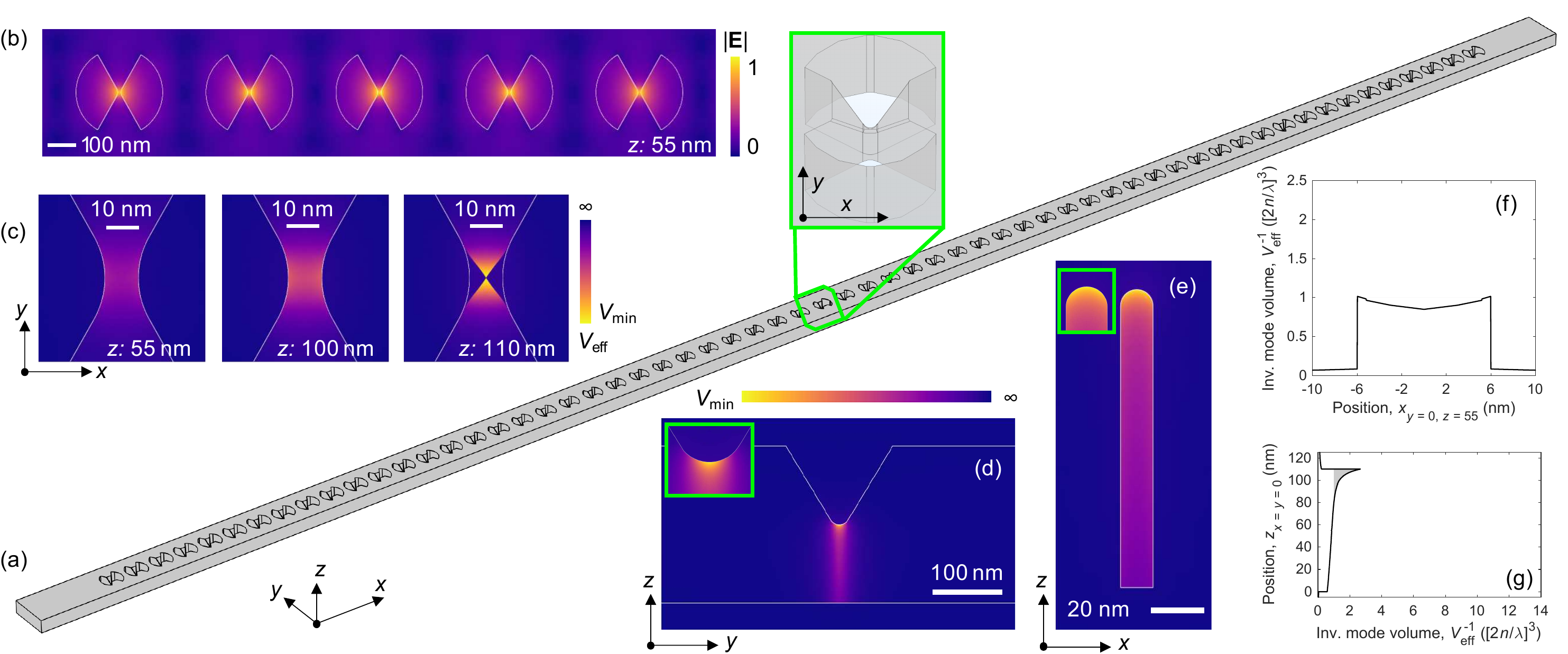}
    \caption{
    Model of a nanocavity with an array of V-groove bowties that have a gap, $g = \SI{12}{nm}$, where lightning-rod effects imply mode volumes below the diffraction limit at the V-groove tip and bulk confinement above the diffraction limit.
    (a), The cavity is a suspended photonic-crystal nanobeam-cavity defined in silicon. The inset shows a V-groove penetrating half way into the bowtie that creates a lightning-rod effect.
    (b-c), Electric field norm, $|\mathbf{E}|$, and effective mode volume, $V$, respectively, in the ($xy$) cross-section plane, shown at the center of the structure, close to the interface, and at the interface.
    (d-e), Side-view of $V$ in $yz$- and $xz$-planes with insets showing the apex of the lightning-rod.
    (f-g), Line-scans through the structure showing $1/V$ normalized to the inverse of the diffraction limit; the gray-shaded area highlights the regions with a mode volume below the diffraction limit.
    }
    \label{fig:5}
\end{figure*}

The example in Fig.~\ref{fig:3} is hypothetical, constructed to illustrate the physically different regimes associated with evaluating mode volumes in the geometric center of nanocavities compared to at the points with the highest field intensity. However, our insights have significant impact also on current experimental research. We now consider the recent experiments on dielectric nanocavities by Hu et al.\ \cite{hu_experimental_2018}, where a photonic-crystal nanobeam cavity with an array of V-groove bowtie unit cells was considered. As we will show, the interplay between the bowtie effect and the lightning rod due to the V-groove has decisive impact on the numerical challenges as well as on the conclusions that can be drawn from the experiment. Due to significant differences between the designed and the actually fabricated cavities in this experiment, we consider both independently. Starting with the designed structure, Fig.~\ref{fig:4}(a) shows the nanobeam cavity design consisting of a central bowtie, 20 tapered unit cells on both sides, and 10 additional bowties of the largest tapered bowtie forming a mirror to obtain a high quality factor; the bowtie has a designed gap, $g=\SI{4}{nm}$. The nanobeam is \SI{220}{nm} thick and \SI{700}{nm} wide. The inset shows the central unit cell where a V-groove is introduced to penetrate half-way (\SI{110}{nm}) through the membrane.
This causes a shift of confinement regime from bulk to lightning-rod as discussed further below.

We estimate an in-plane radius of curvature, $R_{xy}=\SI{30}{nm}$, from the geometry used in the theoretical model by Hu et al.
We find the fundamental mode of the cavity using a finite-element method solving a 3D electrodynamic problem. Figure~\ref{fig:4}(b) shows the norm of the electric field, $|\mathbf{E}|$, mid-way between the tip of the V-groove and the bottom of the structure, i.e., at $z=\SI{55}{nm}$ for the five central unit cells. The mode exhibits an array of hotspots at each bowtie, modulated by an envelope along the tapered unit cells. Since $R_{xy}>g$ and $\theta=\SI{120}{\degree}$ for the in-plane geometry, the confinement would have been in the bulk regime, cf.~Figs.~\ref{fig:2}(c) and (d), in the absence of the V-groove.
Figure~\ref{fig:4}(c) shows the normalized energy density, proportional to the inverse mode volume, at different distances from the V-groove tip of the central bowtie in Fig.~\ref{fig:4}(b), with $z=\SI{55}{nm}$ corresponding to the middle of the structure between the V-groove and the bottom of the cavity shown in Fig.~\ref{fig:4}(b) and $z=\SI{110}{nm}$ corresponding to the tip of the V-groove. These colormaps are normalized to the minimum mode volume of the structure, which occurs at the very tip of the V-groove at $z=\SI{110}{nm}$, and show that the mode is visibly much less enhanced in the bulk only \SI{10}{nm} away from the tip.

Figure~\ref{fig:4}(d-e) shows $1/V$ at two planes cut through the center of the structure in the $yz$- and $xz$-planes, respectively, with insets showing the lightning-rod tip caused by the small radii of curvature estimated from the theoretical geometry, $R_{yz}=\SI{2}{nm}$ and $R_{xz}=g/2=\SI{2}{nm}$. The mode is enhanced in the dielectric material just at the interface and drops rapidly away from the boundary, similarly to the effect observed in Fig.~\ref{fig:1}(d).
Figure~\ref{fig:4}(f) shows $1/V$ across the center at $z=\SI{55}{nm}$, i.e., a horizontal line-scan through the center of Fig~\ref{fig:4}(b). This confirms that the symmetric bowtie effect in the $xy$-plane leads to a slowly varying field inside the material and thus a bulk confinement regime in the absence of the V-groove as already predicted from the geometric parameters.
In contrast, Figure~\ref{fig:4}(g) shows $1/V$ along $z$, i.e., vertically through the centers of Figure~\ref{fig:4}(d) and (e), where the V-groove acts as a lightning rod enhancing the mode volume at the surface, because both $R_{xz}$ and $R_{yz}$ are much smaller than the height of the structure, as shown conceptually in Figs.~\ref{fig:1}(d)-(f). The shaded areas in Figs.~\ref{fig:4}(f) and (g) show regions where the mode volume is below the diffraction limit, and it can be readily seen from Fig.~\ref{fig:4}(g) that the mode volume increases from $<(1/10) [\lambda/(2n)]^3$ to $\sim(1/2) [\lambda/(2n)]^3$ from the lightning-rod tip and into the bulk, i.e., almost an order of magnitude difference. Even though the mode volume stems from the lightning-rod effect, it remains below the diffraction limit inside the structure due to the small designed gap, $g = \SI{4}{nm}$.
More specifically, we calculate the mode volume of this theoretical structure at the tip of the V-groove, $\mathbf{r}_\text{min}=(x=0, y=0, z=110)\ \text{nm}$, to be $V_\text{tip}^\text{t} = 0.073 [\lambda/(2n)]^3$. This is in good agreement with the claimed value, $V_\text{min}^\text{ref} = 0.067 [\lambda/(2n)]^3$. However, the mode volume in the center, i.e., $\mathbf{r}_\text{c}=(x=0, y=0, z=55)\ \text{nm}$, is $V_\text{cen}^\text{t} = 0.51 [\lambda/(2n)]^3$ which is much larger but still below the diffraction limit.

We now consider the same cavity as in Fig.~\ref{fig:4} but for the dimensions which Hu et al. measured on the fabricated structure. The result is shown in Fig.~\ref{fig:5}. All components are identical to Fig.~\ref{fig:4}, except the bowtie gap, which is changed to the value measured by scanning electron microscopy, $g=\SI{12}{nm}$, and we estimate the following radii of curvature from the scanning electron micrographs provided by Hu et al.: $R_{xy}=\SI{30}{nm}$, $R_{yz}=\SI{15}{nm}$, and $R_{xz}=g/2=\SI{6}{nm}$, which we use to build the numerical model shown in Fig.~\ref{fig:5}(a). The resulting mode is shown in Figs.~\ref{fig:5}(b-g) and it is qualitatively similar to that found in Fig.~\ref{fig:4} although with important differences. In particular, the minimum mode volume of this structure, $V_\text{min}^\text{SEM} = 0.36 [\lambda/(2n)]^3$, which occurs at the tip of the fabricated V-groove as shown in Figs.~\ref{fig:5}(d)-(g), is $\sim6$ times larger than for the designed structure due to the larger gap and, importantly, the larger radii of curvature. We find the mode volume at the center of the structure, i.e., the mode volume associated with bulk confinement, $V_\text{cen}^\text{ref} = 1.2 [\lambda/(2n)]^3$, which is above the diffraction limit and comparable to conventional nanocavities such as H1 photonic-crystal cavities \cite{nakayama_effect_2011} with $V \sim 2.6 [\lambda/(2n)]^3$.
The larger radii of curvature also imply that the field falls off slower compared to the designed structure, which for example can be seen in Figs.~\ref{fig:4}(d), (e), and (g). Again, the shaded areas highlights the regions where the mode volume is below the diffraction limit. Our estimates of the radii of curvature are associated with substantial uncertainties but we believe they are conservative and therefore provide lower bounds to the calculated mode volumes.

\section{Discussion and conclusion}

In summary, we have shown that the ratio of the gap to the radius of curvature divides bowtie cavities into two confinement regimes: bulk confinement versus lightning rods at surfaces. An important conclusion is that while smaller gaps, $g$, always lead to smaller mode volumes, there is an optimum value of the radius of curvature, which for the optimal bowtie angle of \SI{90}{\degree} is $R_\text{min}=g$, and it is not advantageous to decrease the radius of curvature below this threshold if the goal is enhancing the light-matter interaction inside the material. For a radius of curvature below $R_\text{min}$, the bowtie operates instead as two opposing lightning rods. This distinction is crucial for research on semiconductor devices such as nanolasers and nonlinear optical devices as well as on electronic-photonic integration, which rely on field enhancements inside the dielectric material. Lightning rods may have interesting applications for enhancing surface interactions, but they do not necessarily imply a strong light-matter interaction inside the material. Since defects or sharp corners on the surface of any dielectric structure can lead to lightning-rod effects, one might wonder why they are not dominating experiments on transmission in waveguides, near-field spectroscopy, and essentially all optical experiments. The answer may be that these surface effects are so localized and broadband that they average out in actual experiments. In this context, it is interesting to note that strongly localized fields also appear at sharp metallic tips, but more advanced hydrodynamic models of the carriers have shown that such surface effects arise from extrapolation of continuum theory beyond its validity \cite{mortensen_generalized_2014}.

We presented detailed numerical models of nanocavities considered in recent experiments to demonstrate the relevance of our new insights. Since the mode volume can only be measured indirectly through measurements of the LDOS \cite{garcia_de_abajo_optical_2010,lodahl_interfacing_2015}, it may be more useful to infer the mode volume from the fabricated geometry. This means in turn that it becomes critical to accurately estimate the radius of curvature of the fabricated structures, such that theoretical predictions of experiments remain accurate and optimized designs perform as intended. The need for careful measurements of structural parameters is of course not new in nanocavity research but there is a main difference between conventional high-$Q$ nanocavities and cavities with mode volumes below the diffraction limit: In the former case, the most important geometry parameters govern structural disorder, which has a decisive impact on $Q$ while leaving the mode volume approximately invariant. For bowtie cavities, however, the gap and the radius of curvature are also essential fabrication constraints and they have a decisive impact on $V$. The interplay between disorder, gap, and radius of curvature and their impact on $Q$ has not been studied and would be an important line of future research. Ideally, the fabrication constraints should be included already at the design stage \cite{wang_maximizing_2018}, which would require an unprecedented integration of nanofabrication, design, and optical experiments.

\section*{Acknowledgments}
We gratefully acknowledge financial support from the Villum Foundation Young Investigator Program (Grant No. 13170), the Danish National Research Foundation (Grant No. DNRF147 - NanoPhoton), and Innovation Fund Denmark (Grant No. 0175-00022 - NEXUS).

\bibliographystyle{apsrev4-2}
\bibliography{main}

\end{document}